\documentstyle[aas2pp4]{article}
\slugcomment{To appear in Astrophysical Journal Letters}

\lefthead{Gallego et al.}
\righthead{SFR in the Local Universe}

\begin{document}

\title{The current Star Formation Rate of the Local Universe}

\author{J. Gallego}
\affil{Departamento de Astrof\'{\i}sica, Universidad Complutense de Madrid
    E-28040 Madrid, Spain \\ E-mail: jgm@ucmast.fis.ucm.es}

\author{J. Zamorano}
\affil{Departamento de Astrof\'{\i}sica, Universidad Complutense de Madrid
    E-28040 Madrid, Spain \\ E-mail: jaz@ucmast.fis.ucm.es}

\author{A. Arag\'{o}n--Salamanca}
\affil{Institute of Astronomy, Madingley Road, Cambridge CB3 0HA, England
\\ E-mail: aas@ast.cam.ac.uk}

\and

\author{M. Rego}
\affil{Departamento de Astrof\'{\i}sica, Universidad Complutense de Madrid
    E-28040 Madrid, Spain \\ E-mail: mrf@ucmast.fis.ucm.es}

\begin{abstract}
The Universidad Complutense de Madrid (UCM) Survey is a long-term
project aiming at finding and analysing star-forming galaxies detected
by their H$\alpha$ emission in Schmidt objective-prism plates.  The
instrumental set-up limits the volume of the Universe surveyed to a
redshift $z\lesssim0.045$.  So far we have discovered several hundred
emission-line galaxies (ELGs) undergoing star formation at a wide range
of levels.  Analyzing a complete sample of ELGs from the UCM survey we
have computed the H$\alpha$ luminosity function for the star-forming
galaxies in the  surveyed volume of the Universe.  A Schechter function
provides a good fit to the H$\alpha$ luminosity function with the
following parameters:  $\alpha = -1.3\pm0.2$; $L^*({\rm H}\alpha) =
10^{42.15\pm0.08}\,$erg$\,$s$^{-1}$ and
$\phi^*=10^{-3.2\pm0.2}\,$Mpc$^{-3}$ for  $H_0 =
50\,$km$\,$s$^{-1}\,$Mpc$^{-1}$.  Integrating over the full range of
luminosities, we obtain an H$\alpha$ luminosity density of
$10^{39.1\pm0.2}\,$erg$\,$s$^{-1}\,$Mpc$^{-3}$.  Using the H$\alpha$
emission as a star formation rate (SFR) estimator, this translates into
a SFR density for the Local Universe of
$0.013^{+0.007}_{-0.005}\,M_{\odot}\,$yr$^{-1}\,$Mpc$^{-3}$ in star-forming
galaxies with $EW($H$\alpha+$[NII]$)>10$\AA\ and $z\lesssim0.045$, for
a Scalo Initial Mass Function.  This is the first observational
determination of this quantity, which will provide a direct test of
current galaxy formation and evolution models.
\end{abstract}


\section{Introduction}
Studying the properties  of star forming galaxies at low redshift, and
quantifying the amount and location of star formation in the Local
Universe is a necessary step towards understanding galaxy formation and
evolution.  It can provide a crucial test for galaxy formation and
evolution models. Such models are now able to put together cosmology,
dark matter, gas and stars, and make definite predictions for the
properties of the present-day galaxy population and its evolution with
redshift.  In particular, they can predict current SFRs (Kauffmann,
White \& Guiderdoni 1993, Cole et al. 1994) but very few observational
constrains exist.  Substantial data sets have been put together (e.g.
Kennicutt 1992, and references therein) but not for a star-formation
selected sample of galaxies.

Moreover, extensive observational effort has been devoted to the study
of the properties and evolution of high redshift galaxies (e.g. deep
counts, deep redshift surveys, gravitational lensing) but one of the
major problems that arises when analysing high-$z$ data is how to make
a meaningful comparison with local samples, usually obtained using very
different observational techniques. Carrying out such a comparison
involves to distinct steps:  first we need well-understood data for
local samples that are directly related to the distant ones, and
second, we have to transform the apparent properties of the local
objects to predict how they would look like at high-$z$ (i.e.,
introduce cosmological corrections and selection effects). If either or
both of these steps is not accurate enough, apparent differences between
local and distant samples could mimic evolution, and lead to erroneous
conclusions.  A clear example where this problem is very important
occurs in the interpretation of the so-called `faint blue galaxies'
problem, where much of the recent controversy  depends very much on what
different authors believe about the properties of local galaxy samples,
and in particular about the local galaxy LF. See, e.g., Koo \& Kron
(1992) and Broadhurst, Ellis \& Glazebrook (1992) for different views
on this problem.

Closely related to this is the recent evidence found in deep galaxy
redshift surveys selected in the optical and in the near infrared which
suggests star-formation activity substantially increases with redshift
from $z\simeq0$ to $z\simeq1$ (Songaila et al. 1994, Colless 1995,
Ellis et al. 1995, Lilly et al. 1995). Although the causes of this
increase in star formation are not yet clear (interactions and mergers
have been suggested), it could explain the faint blue galaxy excess and
the suggested evolution in the optical field LF (see Arag\'on-Salamanca
1995 for a recent review). Paradoxically, we seem to know more about
the numbers of star-forming galaxies at intermediate and high redshifts
than locally. This letter tackles this problem directly,
providing the first published estimate of the numbers of nearby
star-forming galaxies and the star formation rate (SFR) density of the
Local Universe.

The H$\alpha$ luminosity is a very good direct measurement of the
current star formation rate, since it is directly related to the number
of massive stars (see, e.g., Kennicutt 1992, Gallagher and Gibson
1993).  It is better than other optical Balmer lines such as H$\beta$
or H$\gamma$, affected more by stellar absorption and reddening.
Metallic nebular lines like [OIII] $\lambda$5007, [OII] $\lambda$3727
(affected by excitation and metallicity) or IRAS fluxes (affected by
the dust abundance and properties) are rather star-formation {\it
indicators\/} than quantitative measurements (see Gallagher et al.  1989,
Kennicutt, 1992). Although it could be argued that near infrared
recombination lines such as Br$\gamma$ could be even better
measurements of the current SFR, being less affected by extinction than
the optical lines, the lack of near-IR instrumentation with very large
field of view makes large area near-IR surveys impractical at the
moment.  These considerations imply that the best available way to
quantify current star formation in the local Universe is by using an
H$\alpha$-selected sample of galaxies. However, it should be kept in mind that
if some fraction of newly formed stars is completely hidden by dust, they would
not contribute to the H$\alpha$ luminosity, implying that the SFR derived here
should be considered a lower limit.

The Universidad Complutense de Madrid (UCM) Survey for emission-line
objects provides an ideal tool for such purpose, since galaxies were
selected according to their H$\alpha$ emission.  It is a low-dispersion
objective-prism survey for low-redshift emission-line galaxies (ELGs)
being carried out using the 80/120$\,$cm f/3 Schmidt telescope at the
German-Spanish Observatory of Calar Alto (Almer\'{\i}a, Spain).
A $4\arcdeg$ full aperture prism, yielding a dispersion of
$1950\,$\AA/mm, combined with IIIaF emulsion, has been used to search
for ELGs selected by the presence of H$\alpha$ emission in their
spectra.  The instrumental setup is able to record the H$\alpha$+[NII]
$\lambda\lambda6548,6584$ blend in emission for objects up to
$z\lesssim0.045$.
The survey itself and the two first discovery lists
have been described in detail by Zamorano et al. 1994 \& 1995.
Preliminary results and details are presented in Rego et al. (1989,
1993), and Zamorano et al. (1990, 1992).
So far, the UCM Survey has found 264 candidates in 17 fields covering
471.4 square degrees. The overall object density is 0.56 candidates per
square degree.
% , i.e. six times that of the Markarian Survey (Mazarella
% \& Balzano 1986).
More than half of the sample (138 candidates) are galaxies which do not
appear in any published catalogue.  Follow-up imaging and spectroscopy
of the survey candidates has been completed and used to confirm the
candidates as ELGs and classify them.  Morphologically, the UCM sample
represents a heterogeneous population of galaxies dominated by late-type
galaxies (66\% being Sb or later) with about 10\% presenting typical
parameters of earlier types, and another $\simeq10$\% being
irregulars.
In all but 22\%
of the galaxies we find the emission to come almost entirely from the nuclear
regions.
 Their median luminosity is
about one magnitude fainter than that of magnitude-limited galaxy
samples selected in blue plates, with a higher fraction of
low-luminosity galaxies.  As classified from their spectra and IRAS
data, the  most commonly found ELGs (47\%) are intermediate to
low-luminosity objects with a very intense star-formation region which
dominates the energy output of the galaxy.  Their metallicities
range from solar values to ${1\over{40}}Z_{\odot}$, peaking at
${1\over4}Z_{\odot}$.  A full analysis has been published in Vitores et
al. (1995a, 1995b \& 1995c) and Gallego et al.  (1995a, 1995b \&
1995c). This extensive data set provides the raw materials for the
study presented here.

 From this catalogue, we have built a representative complete sample of
star-forming galaxies (which excludes AGNs) suitable for the statistical study
presented in this {\it Letter}. In section~2 we discuss the galaxy selection
and estimate the H$\alpha$ luminosity function, which in section~3 is used to
estimate the present day SFR density.  Finally, we summarise our conclusions
in section~4.

\section{The H$\alpha$ luminosity function}

Direct information on the amount and distribution of the present-day
SFR in the Local Universe can be obtained by constructing the
luminosity function (LF) for galaxies with current star-formation activity.
The Balmer line emission, and in particular the H$\alpha$ luminosity,
provide a direct signature
of the star-formation activity occurring in the galaxies. We will now estimate
the H$\alpha$ LF.

In a sample built from an objective-prism survey such as the UCM one,
the survey completeness depends primarily on the line flux, not the apparent
magnitude. As shown by Salzer (1989), the completeness of a line-selected
sample depends on the emission-line+continuum flux and the contrast between
the line and continuum (the equivalent width of the line). Applying the
procedures of Salzer (1989) to the UCM sample (Gallego 1995c) we
define an arbitrary magnitude scale $m_{{\rm L}+{\rm C}}$ as a function of the
line+continuum flux $F_{{\rm L}+{\rm C}}$ (in erg$\,$s$^{-1}\,$cm$^{-2}$):
\begin{equation}
m_{{\rm L}+{\rm C}} = -17.0 - 2.5 \log (F_{{\rm L}+{\rm C}})
\end{equation}
A quantitative estimate of the survey
completeness can be obtained using the $V/V_{\rm max}$ test
(Schmidt 1968, Huchra and Sargent 1973). Such a test indicates that
all the galaxies in the UCM sample with $m_{{\rm L}+{\rm C}} \leq 17.3$
and $EW$(H$\alpha$+[NII]$) \geq 10$\AA\ have  $V/V_{\rm max}\simeq 0.5$, thus
constituting a complete sample (see Gallego et al. 1995c for details).
For the remainder of this study, the sample of 176 UCM galaxies
that fulfil these criteria
will be referred to as ``the UCM representative complete sample''. The
$m_{{\rm L}+{\rm C}}\leq 17.3$ limit corresponds to a line plus continuum
flux of $1.9 \times 10^{-14}\,$erg$\,$s$^{-1}\,$cm$^{-2}$.

With such a sample  we are able to compute the H$\alpha$ luminosity
function.  The total H$\alpha$ luminosity, reddening-corrected using
the Balmer decrement,  has been taken from  Gallego et al. (1995b).
Distances have been estimated from the galaxy redshifts, assuming a
value of $H_0 = 50\,$km$\,$s$^{-1}\,$Mpc$^{-1}$. Note that when
considering the object detection procedure, the combined
H$\alpha$+[NII] emission has been taken into account, since the
objective-prism produces very low resolution. However, when computing
H$\alpha$ luminosities, the [NII] contribution is not included since
the follow-up spectroscopy separates both lines.

If $\Phi[\log L({\rm H}\alpha)]$ is the number of galaxies per unit volume
per 0.4 interval in $\log L({\rm H}\alpha)$, then the Schmidt estimator
can be written as
\begin{equation}
\Phi[\log L({\rm H}\alpha)] = \frac{4\pi}{\Omega} \sum_i \frac{1}{V_{\rm
max}^i}
\end{equation}
where  $\Omega$ is the surveyed solid angle and $V_{\rm max}^i$ represent the
volume enclosed by a sphere of radius equal to the maximum distance the galaxy
could have and still be detected in the survey.
The summation is over all galaxies in the H$\alpha$ luminosity
range $\log L({\rm H}\alpha) \pm 0.5\Delta \log L({\rm H}\alpha)$.
We have used $\Delta \log L({\rm H}\alpha)=0.4$.
The luminosity function for the UCM galaxies is given in Table \ref{tab-phi}.
\placetable{tab-phi}

The $V_{\rm max}$ method presented here has the advantage of
simplicity, but because galaxies are assumed to be distributed
homogeneously, the results may be distorted at the faint end if there
are local inhomogeneities  in the sample.  The maximum likelihood estimator
described in Mobasher, Sharples \& Ellis (1993) is independent of such
inhomogeneities, provided that the LF has the same shape everywhere.
Using this method, we obtain a LF that agrees, within the errors, with
the one obtained from the $V_{\rm max}$ method, showing that local
inhomogeneities are either non-important or average out in the volume
sampled by the UCM survey.

Typical H$\alpha$ luminosities range from $10^{40}$ to
$10^{42}\,$erg$\,$s$^{-1}$.
The density of star-forming galaxies decreases with
H$\alpha$ luminosity, as pointed out previously by other authors (Salzer 1989,
Boroson et al. 1993).  However, and what is more important, the
H$\alpha$ technique used for the UCM survey recovers a higher number of
star-forming galaxies at lower relative levels of star-formation activity
than surveys carried out in blue region of the spectrum.
This selection effect has clear repercussion when estimating the
number and distribution of star-forming galaxies from such blue surveys, since
they can miss up to $\sim50$\% of the galaxies undergoing star formation
and bias the H$\alpha$ luminosity distribution
(thus the SFR distribution) towards higher luminosities (SFRs).

\section{The SFR density of the Local Universe}
One of the most important applications of the H$\alpha$ luminosity
function is that it can be used to estimate the current ``SFR function'' of
galaxies, which describes the
number of star-forming galaxies as a function of their on-going SFR.
Moreover, if an analytical expression is obtained, it can be used to
extrapolate observed luminosity density to total luminosity density
in the volume considered.
The luminosity function for our complete
sample of current star-forming galaxies is well fitted by a Schechter (1976)
function
\begin{equation}
\phi(L) \: dL = \phi^* \: (L/L^*)^{\alpha} \: \exp (-L/L^*) \: d(L/L^*)
\end{equation}
where the luminosity function $\phi(L)$ is directly related to
$\Phi[\log L({\rm H}\alpha)]\,$ for $L = L({\rm H}\alpha)$ by the equation
\begin{equation}
 \Phi (\log L)  \; {{d\log L}\over{0.4}} = \phi(L) dL
\end{equation}
The best-fitting parameters (excluding the first point) are
\begin{eqnarray}
\alpha   & = & -1.3 \pm 0.2 \nonumber \\
 \phi^* & = & 10^{-3.2 \pm 0.2}\,{\rm Mpc}^{-3} \nonumber \\
L^* & = &  10^{42.15 \pm 0.08}\,{\rm erg}\,{\rm s}^{-1} \nonumber \\ \nonumber
\end{eqnarray}
Figure~1 shows the binned luminosity distribution, and  the
solid curve represents the Schechter function with the above parameters.

The errors associated with the derived parameters were obtained using a
Monte-Carlo method.  A large number of simulations were computed with
errors in the LF following a Gaussian distribution of $\sigma$ equal to
the square root of the number of galaxies in each $\log L({\rm
H}\alpha)$ bin.  A Schechter function was then fitted to the simulated
LFs, yielding a normal distribution for the parameter values. The
quoted errors are $1\sigma$ for these distributions.

\placefigure{fig1}

If a Schechter function with the above parameters
is a good approximation to the luminosity
distribution in the volume of Universe considered,  since $\alpha \leq -2$,
$\phi(L)$ is finite and can be integrated for whole range of luminosities.
\begin{eqnarray}
L_{tot} & = & \int^{\infty}_0 \phi(L) \: L \: dL      \nonumber \\
        & = & \phi^* \: L^* \: \Gamma(2+\alpha)     \\   \nonumber
\end{eqnarray}
The gamma function $\Gamma(2+\alpha)$  takes the value
$1.30$ for $\alpha=1.3$, yielding a total H$\alpha$ luminosity per unit
volume of $10^{39.1\pm0.2}\,$erg$\,$s$^{-1}\,$Mpc$^{-3}$  in the Local
Universe ($z\lesssim0.045$) for star-forming galaxies with $EW({\rm
H}\alpha+[{\rm NII}])>10$\AA.

We can now relate this H$\alpha$ luminosity to the SFR following the
method introduced by Kennicutt (1983). We use the evolutionary models
of Charlot \& Bruzual (1993) to relate the current SFR to the number of
ionising photons produced by the newly formed stars, and then Case B
recombination theory to predict the luminosity of the H$\alpha$
emission line. A Scalo (1986) Initial Mass Function, including stars in the
$0.1M_{\odot}<M<125M_{\odot}$ mass range was used. The transformation can
be written as
\begin{equation}
L({\rm H}\alpha) = 9.40 \times 10^{40} \frac{SFR}{M_{\odot} \:
{\rm yr}^{-1}} \; {\rm erg \: s^{-1},}
\end{equation}
The top axis of Figure~1 shows how the $L({\rm H}\alpha)$ scale
transforms into a SFR scale using this equation, so that the current
``SFR function'' can be directly read from the figure.
The total H$\alpha$ luminosity density translates into a SFR density of
$0.013^{+0.007}_{-0.005}\,M_{\odot}\,$yr$^{-1}\,$Mpc$^{-3}$, where the main
source
of uncertainty originates in the normalization factor $\phi^*$ of the LF.

Some caution is necessary when interpreting these quantitative
results.  The H$\alpha$ luminosity is only sensitive to the SFR in
stars with  $M\ge10M_{\odot}$, the main contributors to the ionising
flux. Thus, the total SFR given here is an {\it extrapolation\/}
assuming a reasonable (local neighbourhood) IMF. To transform the total
SFR values given here into SFRs in stars with $M\ge10M_{\odot}$, a
factor of $0.044$ should be applied. Nevertheless, we have decided to
quote total SFRs since they can provide useful constraints for galaxy
formation and evolution models, notwithstanding the IMF uncertainty.

\section{Summary and Conclusions}
Using an H$\alpha$ selected sample of nearby galaxies we have estimated
the H$\alpha$ luminosity function for the local Universe. We argue
that, since the H$\alpha$ emission provides a good estimate of the
on-going star formation, the galaxies have been selected by their
current SFR. Therefore, we can use the H$\alpha$ luminosity function to
determine the ``SFR function'' describing the number of star-forming
galaxies as a function of their SFR. Integrating over all H$\alpha$
luminosities (or SFRs) we determine the current SFR density of the
local Universe for galaxies with $EW($H$\alpha+$[NII]$)>10$\AA\ and
$z\lesssim0.045$.  The transformation of H$\alpha$ luminosities into SFRs
depends on the stellar initial mass function. The values presented in
Section~3 correspond to a solar neighbourhood Scalo (1986) IMF, but can
be easily obtained for any IMF by changing equation~6.

We stress that this has been done for the fist time for a sample of
star-forming galaxies directly selected, as far as it is possible
today, by their SFR. Using H$\alpha$ as the SFR estimator has clear
advantages over optical surveys carried out in the blue and far
infrared surveys. It provides a cleaner sample, less biased by
metallicity, excitation and dust-properties effects,  and covers a very
broad range in star-formation activity, reaching relatively low
levels.

Both the SFR function and the total SFR density determined here
can provide {\it direct\/} constraints to galaxy formation and
evolution models.  They can also play an important role in studies of
high redshift galaxy samples providing the $z=0$ benchmark to be
compared with the numbers of high-$z$ star-forming objects.

\acknowledgments

We thank S. White, A.G. Vitores, O. Alonso, J. Gorgas, and N. Cardiel  for
their many valuable  comments. This work was supported in part by the Spanish
''Programa Sectorial  de Promoci\'{o}n del Conocimiento'' under grant
PB93--456. AAS acknowledges generous financial support from the Royal Society.

%\clearpage

\begin{deluxetable}{ccc}
\tablewidth{0pc}
\tablecaption{$H\alpha$ Luminosity Function for the UCM Survey}
\tablehead{
\colhead{log L($H\alpha$)} & \colhead{log($\Phi$)} & \colhead{Number of}
\nl
\colhead{[erg s$^{-1}$]} & \colhead{[\# per Mpc$^{3}$ per 0.4} &
\colhead{galaxies}
\nl
\colhead{} & \colhead{interval of log(L(H$\alpha$)]} & \colhead{}}

\startdata
 $40.6$ & $-3.380$ & \makebox[0.5cm][r]{$7$} \nl
 $41.0$ & $-2.878$ & \makebox[0.5cm][r]{$26$} \nl
 $41.4$ & $-3.081$ & \makebox[0.5cm][r]{$43$} \nl
 $41.8$ & $-3.323$ & \makebox[0.5cm][r]{$54$} \nl
 $42.2$ & $-3.654$ & \makebox[0.5cm][r]{$36$} \nl
 $42.6$ & $-4.564$ & \makebox[0.5cm][r]{$10$} \nl
\enddata
\label{tab-phi}
\end{deluxetable}

\clearpage

\figcaption[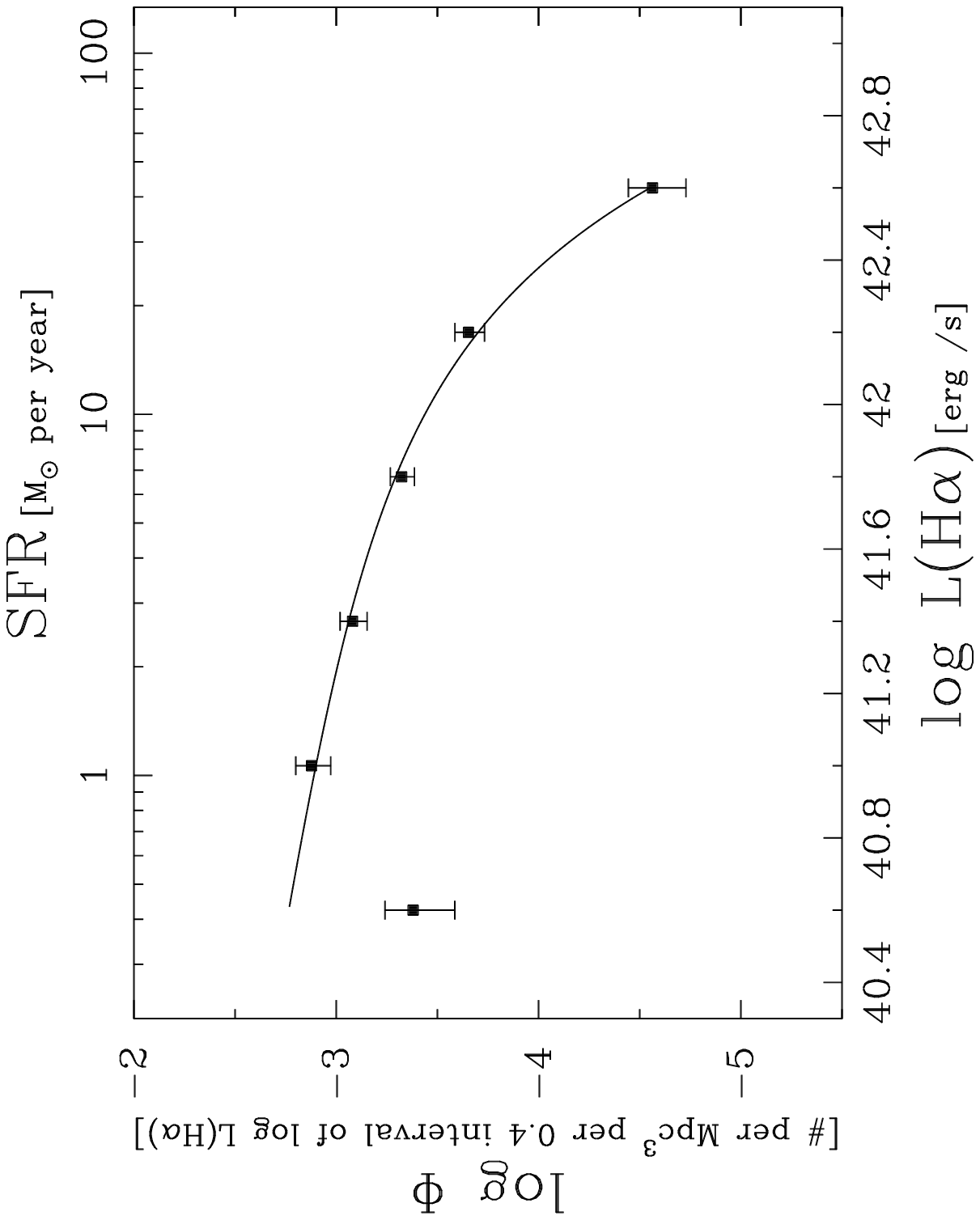]{H$\alpha$ luminosity density distribution
for a complete sample of UCM galaxies. The errors are the square root
of the number of galaxies in each bin.  The solid line represents the
fitted Schechter function to the data points.  The scale at the top of
the diagram indicates the SFR corresponding to the H$\alpha$ luminosity
(see equation~6).\label{fig1}}

\plotone{apjl_fig1_n.ps}
\end{document}